 \documentclass[prl,twocolumn,showpacs]{revtex4}

 \usepackage{graphicx}
 \usepackage{mathtext}
\begin{document}

 \def\BE{\begin{equation}}
 \def\EE{\end{equation}}
 \def\BA{\begin{array}}
 \def\EA{\end{array}}
 \def\ds{\displaystyle}
 \def\BEA{\begin{eqnarray}}
 \def\EEA{\end{eqnarray}}
 \def\r{\vec \rho}
 \def\q{\vec q}
 \def\om{\omega}
 \def\Om{\Omega}
 \def\sinc{{\,\rm sinc}}

\title{Quantum memory for images -- a quantum hologram}

 \author{Denis~V.~Vasilyev$^{1}$, Ivan~V.~Sokolov$^{1,*}$,
and Eugene~S.~Polzik$^{2,3,\dag}$}
 \address{1 V.~A.~Fock Physics Institute, St.~Petersburg University, 198504 Petrodvorets,
  St.~Petersburg, Russia}
 \address{2 QUANTOP, Danish Research Foundation Center for Quantum Optics,
        DK 2100 Copenhagen, Denmark,}
 \address{3 Niels Bohr Institute, DK 2100 Copenhagen, Denmark}

 \begin{abstract}
Matter-light quantum interface and quantum memory for light are
important ingredients of quantum information protocols, such as
quantum networks, distributed quantum computation, etc
\cite{QIPC}. In this Letter we present a spatially multimode
scheme for quantum memory for light, which we call {\sl a quantum
hologram}. Our approach uses a multi-atom ensemble which has been
shown to be efficient for a single spatial mode quantum memory.
Due to the multi-atom nature of the ensemble it is capable of
storing many spatial modes, a feature critical for the present
proposal. A quantum hologram has a higher storage capacity
compared to a classical hologram, and is capable of storing
quantum features of an image, such as multimode superposition and
entangled quantum states, something that a standard hologram is
unable to achieve. Due to optical parallelism, the information
capacity of the quantum hologram will obviously exceed that of a
single-mode scheme.
 \end{abstract}

 \pacs{03.67.Mn, 32.80.Qk}
 %\centerline{Version: \today,}

 \maketitle

One of the challenges in the field of quantum information is the
development of a quantum interface between light and matter. At
the quantum interface quantum states are either transferred
between light and matter (quantum memory) or/and an entangled
matter-light state is generated, which, e.g., is the basis for
quantum teleportation. The quantum memory for light allows for the
high fidelity exchange (transfer, storage and readout) of quantum
states between light and long-lived matter degrees of freedom.
Such interfaces will be an essential component of long distance
quantum communication (quantum repeaters) and quantum computing
networks. Various approaches to the quantum interface with atomic
ensembles have been developed recently, including the
quantum-nondemolition (QND)interaction (for reviews see
\cite{Kuzmich03} and \cite{Advances06}), electromagnetically
induced transparency \cite{Lukin}, and Raman processes
\cite{Kimble,Kuzmich}. The present multi-mode proposal is based on
the QND-type interaction which has been recently used for
high-fidelity quantum memory \cite{Julsgaard04} and teleportation
\cite{Sherson06} of a single-mode light. Up to now the work on
light-atoms interface has been limited to the case of a single
spatial mode of light and a single spatial mode of atomic
ensembles.

On the other hand, multimode parallel quantum protocols for light
only, such as quantum holographic teleportation
\cite{Sokolov01,Gatti04} and quantum dense coding of optical
images \cite{Golubev06} have been elaborated recently. The
protocols of quantum imaging are based on the use of broadband
spatially multimode light beams in an entangled  Einstein -
Podolsky - Rosen (EPR) quantum state.

In this Letter we develop theoretically a multimode parallel
quantum memory for light, where an input signal is carried by a
distributed in space and time wavefront (an optical image). Atomic
ensembles used so far only for a single mode storage are
inherently suitable for quantum holograms due to the possibility
for storage of many spatial modes, which markedly distinguishes
them from a single atom memory.

We utilize the two-pass storage and readout protocols previously
proposed for a single mode scenario \cite{Kuzmich03}. Multimode
generalization for other single mode memory protocols, such as QND
interaction followed by a quantum feedback onto atoms
\cite{Julsgaard04}, and multi-pass protocols
\cite{Hammerer,Sorensen} will be discussed elsewhere.

The key parameter for a quantum hologram is a spatial element
(pixel). We use squeezed light to enhance the fidelity of the
readout of the hologram, therefore in the following we investigate
the relation between the pixel size and the transverse coherence
length of squeezing. We conclude with calculations of the overall
fidelity per pixel for the full cycle of the holographic memory.

\begin{figure}
 \begin{center}
 \includegraphics[width=60mm]{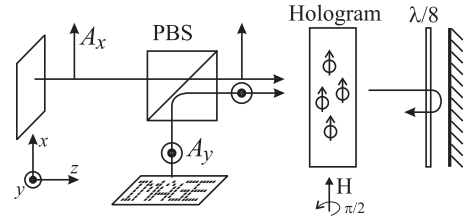}
 \caption{The scheme of the write stage of the quantum hologram}
 \label{fig1_scheme}
 \end{center}
 \end{figure}

The scheme, illustrating the write stage of our quantum memory
protocol is shown in Fig.~\ref{fig1_scheme}.

Consider an ensemble of atoms fixed at random positions with a
spin $\frac{1}{2}$ both in the ground and in the excited state.
The long-lived ground state spin of an atom ${\vec J}^a$ is
initially oriented in the vertical direction $x$. A classical
off-resonant $x$-polarized plane wave of frequency $\om_0$ with a
slowly varying amplitude $A_x$ (taken as real) propagates in the
$z$-direction. An input signal is represented by a weak quantized
$y$-polarized field of the same frequency and average direction of
propagation with an amplitude $A_y(\vec r,t)\ll A_x$. In what
follows we consider this multimode input field in the paraxial
approximation. The QND light-matter interaction leads to two basic
effects: (i) the Faraday rotation of light polarization due to
longitudinal $z$-component of collective atomic spin; and (ii) the
atomic spin rotation, caused by unequal light shifts of the ground
state sub-levels with $m_z=\pm 1/2$ in the presence of circular
light polarization. The relevant part of the Hamiltonian is
\cite{Advances06}:
  \BE
H = \frac{2\pi k_0 |d|^2}{\om_{eg}-\om_0}\int_V d\vec r \sum_a
J_z^a S_z(\vec r,t)\delta(\vec r -\vec r_a).
 \label{hamiltonian}
  \EE
Here $\om_{eg}$ is the frequency of the atomic transition, $d$ is
the dipole matrix element, and $k_0 =\om_0 c$. The $z$-component
of the Stokes vector is $S_z(\vec r,t) = -iA_x\left(A_y(\vec
r,t)-A_y^\dag(\vec r,t)\right)$. The amplitude $A_y$ is defined
via
 $$
 A_y(z,\r,t)=\int\frac{dk_z}{2\pi}\int\frac{d\vec
q}{(2\pi)^2}\sqrt\frac{\omega(k)}{k_0}\,a_y(\vec k)\times
 $$
 \vspace{-6mm}
 \BE
 \exp[i(\q\cdot\r+(k_z-k_0)z-(\omega(k)-\omega_0)t)],
 \label{A_y}
 \EE
here $a_y(\vec k)$ and $a_y^\dagger(\vec k)$ -- the annihilation
and creation operators for the wave $\vec k$, which obey standard
commutation relations $ [a_y(\vec k),a_y^\dagger(\vec
k\,')]=(2\pi)^3\delta(\vec k-\vec k\,')$, $[a_y(\vec  k),a_y(\vec
k\,')]=0$. In the paraxial approximation we have
$[A_y(z,\r,t),A_y^\dag(z,\r\,',t')] =
\delta(t-t')\delta(\r-\r\,')$, where $\vec r=(\r,z)$, $\vec
k=(\q,k_z)$. The space-dependent canonical variables for the input
light are defined as quantities averaged over the interaction time
$T$:
 \BEA
 X_L(\r)&=&
\frac{1}{\sqrt{T}}\int_T dt
\frac{A_y(0,\r,t)+{A_y}^\dag(0,\r,t)}{\sqrt{2}},\nonumber \\
 P_L(\r)&=&
\frac{1}{\sqrt{T}}\int_T dt
\frac{A_y(0,\r,t)-{A_y}^\dag(0,\r,t)}{i\sqrt{2}},
  \label{canonical_light}
 \EEA
and obey the commutation relations
 \BE
 \overline{[X_L(\r),P_L(\r\,')]} =  i \delta(\r-\r\,').
 \label{canonical_comm}
 \EE
In this Letter we will neglect the diffraction over the length $L$
of the atomic layer, thus assuming that the Rayleigh length
associated with the pixel linear size $\sqrt{S}$ is much larger
than $L$, i.~e. $S\gg L\lambda$. A more general theory for quantum
holograms, including the effects of diffraction and spatial
density fluctuations of atoms will be presented in a forthcoming
publication \cite{Vasilyev07}. For a thin atomic layer located at
$z=0$, we introduce the surface density of the collective spin
$\vec J(\r) = \sum_a \vec J^a \delta(\r-\r_a)$. The averaged over
random positions of atoms commutation relation for $y$, $z$
components of the collective spin is
 $$
\overline{[J_y(\r),J_z(\r\,')]} = i\sum_a \langle J_x^a \rangle
\overline{\delta(\r-\r_a)\delta(\r\,'-\r_a)}^a =
 $$
\vspace{-6mm}
 \BE
i n_a \langle J_x^a \rangle \delta(\r-\r\,').
 \label{comm_atoms}
 \EE
Here $n_a$ is the average surface density of atoms. The canonical
variables for the spin subsystem
 \BE
X_A(\r)= \frac{J_y(\r)}{\sqrt{n_a \langle J_x^a \rangle}},\qquad
P_A(\r)=\frac{J_z(\r)}{\sqrt{n_a \langle J_x^a \rangle}},
 \label{canonical_atoms}
 \EE
obey the canonical commutation relations analogous to
(\ref{canonical_comm}).

In what follows both the write and readout procedures are
performed in three steps. The input, two intermediate, and output
variables are labelled by the (in), (1), (2), and (out)
superscripts. The label W (R) indicates the write (readout) stages
of the overall protocol. The transformation of atomic and field
variables in the first passage of the signal looks similar to that
described in \cite{Julsgaard04,Kuzmich03}:
 \BEA
 {X_L^W}^{(1)}(\r) &=& {X_L^W}^{(in)}(\r) + \kappa{P_A^W}^{(in)}(\r),\nonumber\\
 {P_L^W}^{(1)}(\r) &=& {P_L^W}^{(in)}(\r),\nonumber\\
 {X_A^W}^{(1)}(\r) &=& {X_A^W}^{(in)}(\r) + \kappa {P_L^W}^{(in)}(\r)\left(1+
 \frac{\delta J^a_x(\r)}{n_a\langle J^a_x\rangle}\right),\nonumber\\
 \label{first_pass}
 {P_A^W}^{(1)}(\r) &=& {P_A^W}^{(in)}(\r).
 \EEA
Here the coupling constant $\kappa=\alpha_0 \eta=1$, where
$\alpha_0$ is the resonant optical depth and $\eta$ is the
probability of spontaneous emission \cite{Advances06}. Since
$\eta\ll 1$ is required in order to neglect spontaneous emission,
the usual condition $\alpha_0=\lambda^2 n_a/2\pi\gg 1$ should be
fulfilled.

A non-trivial last term in the 3rd equation arises due to spatial
fluctuations of the atomic density. It accounts for the fact, that
the local value of the rotated collective spin and the local value
of the coupling constant may differ from the average value. The
effect of this term  depends on the size of an elementary pixel.
Under the conditions $\lambda^2 n_a/2\pi\gg 1$ and $S\gg
L\lambda$, where $L \gg \lambda$, the number of atoms per pixel is
large, and we can neglect the influence of the atomic density
fluctuations.

At the second step, the atomic spins are rotated around the
$x$-axis by the $\pi/2$ pulse of an auxiliary magnetic field. The
similar rotation of the Stokes vector of light is performed by the
reflection of the signal wave from a mirror and by the double
passage through  $\lambda/8$ plate. This is described by the
transformation ${X_A^W}^{(2)} = - {P_A^W}^{(1)}$, ${P_A^W}^{(2)} =
{X_A^W}^{(1)}$,  ${X_L^W}^{(2)} = - {P_L^W}^{(1)}$, ${P_L^W}^{(2)}
=  {X_L^W}^{(1)}$.

At the third step, the signal wave again propagates through the
atoms. The transformation ``(2) $\to$ (out)'' of the light and
matter variables is the same as at the ``(in) $\to$ (1)'' step,
see (\ref{first_pass}). After all three steps of the write
procedure, we arrive at
 \BEA
 {X_L^W}^{(out)}(\r) &=& {X_A^W}^{(in)}(\r),\nonumber\\
 {P_L^W}^{(out)}(\r) &=& {P_A^W}^{(in)}(\r)+ {X_L^W}^{(in)}(\r),\nonumber\\
 \label{write}
 {X_A^W}^{(out)}(\r) &=& {X_L^W}^{(in)}(\r),\\
 {P_A^W}^{(out)}(\r) &=& {P_L^W}^{(in)}(\r)+ {X_A^W}^{(in)}(\r).\nonumber
   \EEA
As seen from (\ref{write}), the write procedure transfers the
input signal variables onto the  collective atomic spin. One can
achieve a perfect light--matter state transfer provided the
initial fluctuations ${X_A^W}^{(in)}(\r)$ of the collective spin
are suppressed (squeezed) with a sufficient spatial resolution.
Spin squeezing for a spatially single-mode configuration was
demonstrated in \cite{Kuzmich99,Mabuchi}. An extention to a
multi-mode case will be analyzed elsewhere \cite{Vasilyev07}.

Note that after the first pass (\ref{first_pass}) only one
quadrature of light is written onto the hologram. For a classical
hologram this leads to a well known effect when the readout
produces two images: the real and the imaginary one.

The transformation (\ref{write}) describes the state exchange
between light and matter. The same 3-step procedure can be used to
transfer the quantum state of atoms created at the write stage
onto the readout light wave. By substituting in (\ref{write}) the
label W to R, we obtain the transformation for the readout part of
the protocol. The same reasoning as above suggests that for the
high fidelity readout one needs to use the spatially multimode
squeezed light with suppressed fluctuations ${X_L^R}^{(in)}(\r)$.

The ultimate goal of the protocol is to transfer a quantum state
of the input (at the write stage) light signal to the output (at
the readout stage) light. By combining  the described above
transformations, we can relate the input and output variables of
the total write + readout protocol of quantum hologram:
 \BEA
 {X_L^R}^{(out)}(\r) &=& {X_L^W}^{(in)}(\r)+F_X(\r),\nonumber \\
 {P_L^R}^{(out)}(\r) &=& {P_L^W}^{(in)}(\r)+F_P(\r).
  \label{memory}
 \EEA
This transformation is analogous to the one describing quantum
holographic teleportation of an optical image
\cite{Sokolov01,Gatti04}. The noise contributions specific for our
model of memory are given by
 \BE
 F_X(\r)= 0,\qquad
 F_P(\r) = {X_A^W}^{(in)}(\r) + {X_L^R}^{(in)}(\r).
 \label{noise}
 \EE
Consider the field amplitude averaged over the surface $S_i$ of a
square pixel $i$ of area $S$. The averaged noise amplitudes and
the covariance matrix are
 \BE
 F_{X,P}(i) =
 \frac{1}{\sqrt{S}}\int_{S_i}d\r\, F_{X,P}(\r),
 \label{noise_pix}
 \EE
$ C^X(i,j)= \langle F_X(i)F_X(j)\rangle$, $C^P(i,j) = \langle
F_P(i)F_P(j)\rangle$. Assume the input signal field to be in the
spatially multimode coherent state. The fidelity for an array of
$N$ pixels is related \cite{Gatti04} to the covariance matrix as
 \BE
F_N = \left[{\rm det}\big(\delta_{ij} + C^X(i,j)\big)\,{\rm
det}\big(\delta_{ij} + C^P(i,j)\big)\right]^{-1/2}.
 \label{fidelity_matrix}
 \EE

We evaluate the fidelity for two initial states of the collective
spin subsystem: (i) the coherent spin state with atomic spins
oriented in the vertical $x$--direction with the fluctuations
${X_A^W}^{(in)}(\r) = X_A^{(vac)}(\r)$, ${P_A^W}^{(in)}(\r) =
P_A^{(vac)}(\r)$, and (ii) the perfect spin squeezed state with
the same average orientation, when ${X_A^W}^{(in)}(\r) =
X_A^{(sq)}(\r) \to 0$.

The vacuum state quadrature amplitudes, averaged over the pixel,
have the variance  $\langle {X_A}^{(vac)}(i){X_A}^{(vac)}(j)
\rangle = \delta_{i,j}/2$, and similar for ${P_A}^{(vac)}(i)$.

The state of the input light wave used for the readout of the
quantum memory is a spatially multimode squeezed state, $
{X_L^R}^{(in)}(\r) = X_L^{(sq)}(\r)$, ${P_L^R}^{(in)}(\r )=
P_L^{(sq)}(\r)$.

The spatially multimode squeezed light can be generated in a
nonlinear crystal with $\chi^{(2)}$ nonlinearity. For definiteness
we assume the collinear degenerate wave matching in the crystal.
The increase of fidelity is achieved by the suppression
(squeezing) of the quadrature amplitude $X_L^{(sq)}(\r)$. The
squeezing has also a negative effect on the fidelity: the
amplification (anti-squeezing) of the quadrature amplitude
$P_L^{(sq)}(\r)$, followed by scattering on the atomic density
fluctuations. For a moderate squeezing, the relevant contribution
to the noise covariance matrix is estimated \cite{Vasilyev07} as
$C^X(i,i) \sim \exp (2r)/n_a l \lambda$, where $\exp (r)$ is the
amplitude squeezing and $l$ is the parametric crystal length. For
a sufficiently large atomic density this contribution is
negligible.

Consider the contribution to the covariance matrix element
$C^P(i,j) = \langle F_P^\dag(i)F_P(j)\rangle$ coming from squeezed
light:
 $$
{C^P}^{(sq)}(i,j) = \langle
{X_L}^{(sq)\dag}(i){X_L}^{(sq)}(j)\rangle =
 $$
 \vspace{-6mm}
 \BE
\frac{1}{ST}\int_{S_i,S_j} d\r\,' d\r\,''\int_T dt' dt'' \,\langle
{X_L^{(sq)}}^\dag(\r\,',t')X_L^{(sq)}(\r\,'',t'')\rangle.
\label{CP_squeezed}
 \EE
The covariance matrix $\langle
{X_L}^{(sq)\dag}(i){X_L}^{(sq)}(j)\rangle$ of the squeezed light
quadrature components averaged over the observation volume (the
pixel area and the sampling time) determines the noise, the
fidelity, the information capacity, etc. for optical schemes,
considered earlier for optical images: the homodyne detection
\cite{Kolobov89,Kolobov99}, the quantum teleportation
\cite{Sokolov01,Gatti04} and the telecloning. In analogy to
\cite{Gatti04}, we arrive at:
 \BE
{C^P}^{(sq)}(i,j) = \frac{1}{2} \int dq
B_\Delta(\q)\cos[\q(\r_i-\r_j)] G_X(\q,0).
 \label{CP_Green}
 \EE
Here $G_X(\q,\Om)$ is the Green function of the squeezed
quadrature in the Fourier domain,
 $$
\langle X_L^\dag(\q,\Om) X_L(\q\,',\Om')\rangle =
 $$
 \vspace{-9mm}
 \BE
(2\pi)^3\delta(\q-\q\,')\delta(\Om-\Om')\cdot
\frac{1}{2}G_X(\q,\Om),
 \EE
 \label{Green_def}
and $B_\Delta(\q)$ is the delta--like even weight function, which
originates from the integrals over the pixel surface.

Since the interaction time T is much longer than the coherence
time of the squeezed light, only the low frequencies $\Om \to 0$
contribute to (\ref{CP_Green}). In terms of commonly used
parameters of the wide--band squeezing, the Green function is
 \BE
 G_X(\vec q) = e^{2r(\q,\Om)}\cos^2\psi(\q,\Om) +
 e^{-2r(\q,\Om)}\sin^2\psi(\q,\Om).
 \label{Green_r_psi}
 \EE
Here $\exp [-r(\q,\Omega)]$ is the squeezing factor, and
$\psi(\q,\Omega)$ is the orientation angle of the anti-squeezed
axis of the uncertainty ellipse for a given frequency
\cite{Kolobov89,Kolobov99}.

\begin{figure}[h]
 \begin{center}
 \includegraphics[width=60mm]{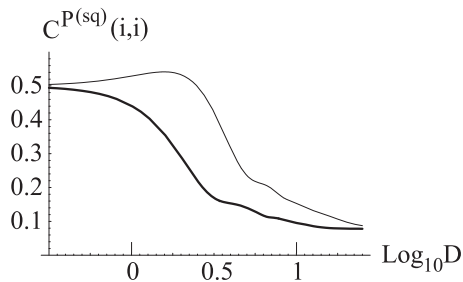}
 \end{center}
\caption{Covariance matrix diagonal element for one pixel (the
bold and hair lines -- with and without phase correction of
squeezing by means of a thin lens). Here $\psi(0,0)=\pi/2$,
$\exp[r(0,0)]=3$.}
 \label{fig2_corrmatrix}
 \end{figure}

For a single pixel the overall fidelity $F_1$ of the
write--readout cycle of the quantum hologram is determined by the
diagonal matrix elements: $C^X(i,i) = 0$, $ C^P(i,i) = 1/2 +
{C^P}^{(sq)}(i,i)$ or $ C^P(i,i) = {C^P}^{(sq)}(i,i)$ for the
coherent and the squeezed initial state of the atomic spin,
respectively.
%%%%%%%%%%%%%%%%%%%%%%%%%%%%%%%%%%%%%%%%%%%%%%%%%%%%%%%%%%%%%%%%
As seen from (\ref{fidelity_matrix}), the ultimate value of
fidelity $F_1 = 1$  is reached for zero diagonal elements of the
noise covariance matrix.
%%%%%%%%%%%%%%%%%%%%%%%%%%%%%%%%%%%%%%%%%%%%%%%%%%%%%%%%%%%%%%%%
In Fig.~\ref{fig2_corrmatrix}
we plot the variance ${C^P}^{(sq)}(i,i)$  for one pixel as a
function of the pixel size $\Delta=\sqrt{S}$, normalized to the
transverse coherence length $l_d$ of the spatially multimode
squeezed light. The latter scale is due to the diffraction of the
downconversion light inside the nonlinear crystal when the
propagation length is of the order of the length of parametric
amplification. For a moderate squeezing a fair estimate for the
coherence length is $l_d \sim \sqrt{l/2k_c}$ (here $k_c$ is the
downconversion wave vector inside the crystal). In our plots
$D=\Delta/l_d$. A properly inserted thin lens is able to
compensate the spatial frequency dispersion (rotation in the plane
of quadrature components) of squeezing ellipses
\cite{Kolobov89,Kolobov99}. The effect of compensation is also
shown in our plots.

As shown in \cite{Gatti04}, the fidelity of the quantum state
transfer for simple multipixel arrays scales approximately as the
$N$-th power of the average fidelity per pixel, $ F_{av} =
(F_N)^{1/N}$. In order to find $F_{av}$ for a large array, the
covariance matrix is transformed to the diagonal form. The
eigenvectors of the matrix $\langle
{X_L}^{(sq)\dag}(i){X_L}^{(sq)}(j)\rangle$ are given by the
superpositions of amplitudes ${X_L}^{(sq)}$ with a quantized 2D
``wave vector''.

\begin{figure}
 \begin{center}
 \includegraphics[width=60mm]{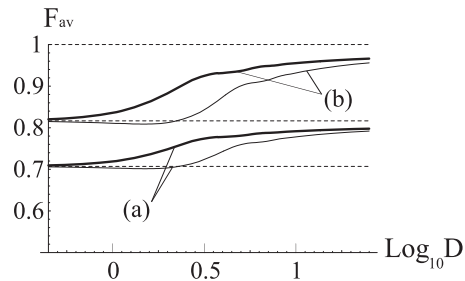}
 \caption{Average fidelity per pixel for the initial coherent state (a) and the
 perfect squeezed state (b) of the collective atomic spin. The
 parameters of light squeezing are the same as in Fig.~\ref{fig2_corrmatrix}.}
 \label{fig3_av_fidelity}
 \end{center}
 \end{figure}

The average fidelity per pixel for our model of quantum memory is
plotted in Fig.~\ref{fig3_av_fidelity}.
For the coherent initial state of the atoms (curves (a)), the
upper limit $F_{av} = \sqrt{2/3} = 0.82$ can be reached for a
large pixel size, $\sqrt{S}\gg l_d$, and perfect squeezing of
light. The fidelity is limited by the vacuum fluctuations of the
initial collective spin. For a small pixel, $\sqrt{S}\ll l_d$, the
light squeezing has no effect. The lower limit $F_{av} =
\sqrt{1/2} = 0.71$ corresponds to the vacuum noise of the initial
state of both atoms and the light used for the readout of quantum
memory. When both the collective atomic spin and the readout light
are prepared in a perfect squeezed state (curves (b)), the perfect
fidelity, $F_{av}=1$, can be achieved for a large pixel size.  The
lower limit $F_{av} = \sqrt{2/3} = 0.82$ is due to the fact, that
for a small pixel size the light fluctuations are restored back to
the vacuum value. The quantum hologram hence provides the fidelity
much higher than the best classical fidelity for the complete
write plus readout protocol which according to \cite{Benchmark} is
$0.5$.

To conclude, we have proposed an essentially parallel model of the
quantum memory for light and analyzed its characteristic spatial
scales, which in our scheme are associated with the transverse
scales of the non-classical light used for the readout of the
quantum hologram.

This research was supported by the RFBR -- CNRS under Project
05-02-19646, by the Ministry of Science and Education of RF under
Project RNP.2.1.1.362, by the INTAS under Project ``Advanced
Quantum Imaging and Quantum Information with Continuous
Variables'', and by the EU projects QAP and COVAQIAL.

\bibliographystyle{plain}

\end{document}